\documentclass[conference,10pt]{IEEEtran}
\usepackage{transparent,comment}
\usepackage[T1]{fontenc}
\usepackage{cite}
\usepackage{multirow}
\usepackage{graphicx}
\usepackage{amssymb,amsmath,times}
\usepackage{pb-diagram}
\usepackage{psfrag,balance}
\usepackage{srcltx,verbatim} 
\usepackage{xcolor}
\usepackage{algorithmic}
\usepackage[ruled,vlined,linesnumbered]{algorithm2e}
\usepackage[caption=false]{subfig} 
\usepackage{array}
\usepackage{booktabs}
\usepackage{enumitem}
\DeclareMathAlphabet\mathbfcal{OMS}{cmsy}{b}{n}
\newcommand\blfootnote[1]{%
}

\usepackage{orcidlink}

\newcommand{\real}[1]{\Re\left({#1}\right)}

\begin{document}
\bstctlcite{IEEEexample:BSTcontrol}

\title{\huge Interference-Aware Downlink Power Control for High-Altitude Platform Stations  with QoS and Max-Min Fairness}

\author{\IEEEauthorblockN
{Rajan Shrestha and Hayder Al-Hraishawi}
\IEEEauthorblockA{\textit{Department of Electrical and Computer Engineering, University of South Florida, Tampa, FL, USA}\\
Email:\{rajanshrestha, hayder\}@usf.edu}
}

\maketitle

\begin{abstract}
High-altitude platform stations (HAPS) are a promising component of non-terrestrial networks (NTNs) for wide-area connectivity. However, serving multiple users over large user footprints requires efficient power-control mechanisms that account for the unique propagation characteristics and strict power constraints of HAPS. This paper develops a tractable analytical and optimization framework for downlink power allocation in HAPS systems under statistical channel state information, 3GPP NTN path loss, and Rician fading. A closed-form expression for the signal-to-interference-plus-noise ratio (SINR) with mean-channel maximal-ratio transmission (MRT) precoding is derived, capturing geometry-dependent inter-user interference coupling. Based on this, two power-control schemes are proposed: quality-of-service (QoS)-constrained power minimization and max--min SINR fairness. The QoS solution is obtained in closed form or through a low-complexity fixed-point iteration, while the max--min solution is obtained through bisection. Numerical results demonstrate up to \(6~\mathrm{dB}\) reduction in required transmit power at moderate-to-high target rates and improved worst-user SINR compared to the uniform power, inverse path-loss, and geometry-aware heuristic allocations. The proposed QoS-optimal scheme also achieves a QoS feasibility probability of approximately \(0.95\) at \(R^{\mathrm{req}}=2.5\) bits/s/Hz and \(P_{\max}=35\) dBm, while the uniform power, inverse path-loss, and geometry-aware heuristic allocations achieve less than \(0.4\). Moreover, the proposed schemes achieve performance comparable to the linear programming (LP)-based benchmarks at a fraction of the computational cost, reducing the average computation time by approximately \(99.97\%\) when \(K=20\).
\end{abstract}


\section{Introduction}\label{sec:introduction}
High-altitude platform stations (HAPS) are emerging as a key component of integrated terrestrial and non-terrestrial networks (NTNs), providing a complementary connectivity layer between terrestrial infrastructure and satellite systems. Operating at stratospheric altitudes of 17--22~km, HAPS combine wide-area and persistent coverage with favorable line-of-sight (LoS) propagation and near-terrestrial latency \cite{Karabulut2021}. These characteristics make HAPS attractive for extending connectivity to remote and underserved regions and for rapidly augmenting terrestrial infrastructure when conventional coverage is limited or unavailable \cite{Hayder_NGSO}. Serving many users across such wide coverage areas requires efficient spatial multiplexing and interference management \cite{Jang2026}. Equipping HAPS with massive MIMO arrays provides the spatial degrees of freedom needed for high spectral efficiency and fine-grained beamforming \cite{Xi2016}. 


In such systems, downlink power allocation directly influences user performance, fairness, and power efficiency. Although HAPS rely on solar energy, their operation remains tightly power-budgeted due to limited nighttime battery capacity and stringent constraints on payload weight, amplifier output power, and thermal dissipation \cite{Farooqi2025}. Therefore, efficient power allocation is  crucial for sustaining service availability,  supporting long-duration operation, and enabling practical large-scale HAPS deployments \cite{Khennoufa2026}.

Existing works on HAPS massive MIMO \cite{Xi2016,Lian2019,Zong2019,Tashiro2023} mainly address capacity, user grouping, beamforming, precoding, and interference management, while power allocation has also been investigated within MIMO-NOMA and joint transmission designs \cite{Ji2021,Shafie2022,Abbasi2024a,Javed2025}. However, the strong LoS propagation in HAPS massive MIMO creates pronounced geometry-dependent inter-user interference, where spatial coupling is governed by user locations, angular separation, and array responses \cite{3gppTR38811V1540}. The impact of these issues on downlink power allocation, particularly on QoS feasibility, minimum required power, and user fairness, has received limited attention. 

To address this gap, we develop geometry-aware downlink power-control schemes for HAPS massive MIMO under statistical channel state information (CSI), Rician fading, and 3GPP NTN channel models~\cite{3gppTR38811V1540}. The proposed framework accounts for geometry-dependent inter-user interference under mean-channel maximal-ratio transmission (MRT) rather than relying solely on large-scale channel gains. The main contributions of this paper are summarized below:
\begin{itemize}
    \item Derive a closed-form SINR expression for HAPS massive MIMO under statistical CSI and 3GPP NTN propagation,  capturing geometry-dependent inter-user coupling.
    
    \item Formulate the QoS-constrained power-minimization problem, derive its feasibility condition and minimum-power solution, and develop a low-complexity fixed-point iteration to obtain the optimal power allocation.

    \item Formulate the max--min fairness problem to maximize the minimum user SINR and solve it via a bisection-based method.
    
    \item Demonstrate significant gains in power efficiency, QoS feasibility, and fairness over conventional allocations, while achieving performance comparable to linear programming (LP)-based benchmarks at substantially lower computational complexity.
\end{itemize}

\begin{figure}[!t]\centering
      \footnotesize 
      \def\svgwidth{200pt}
    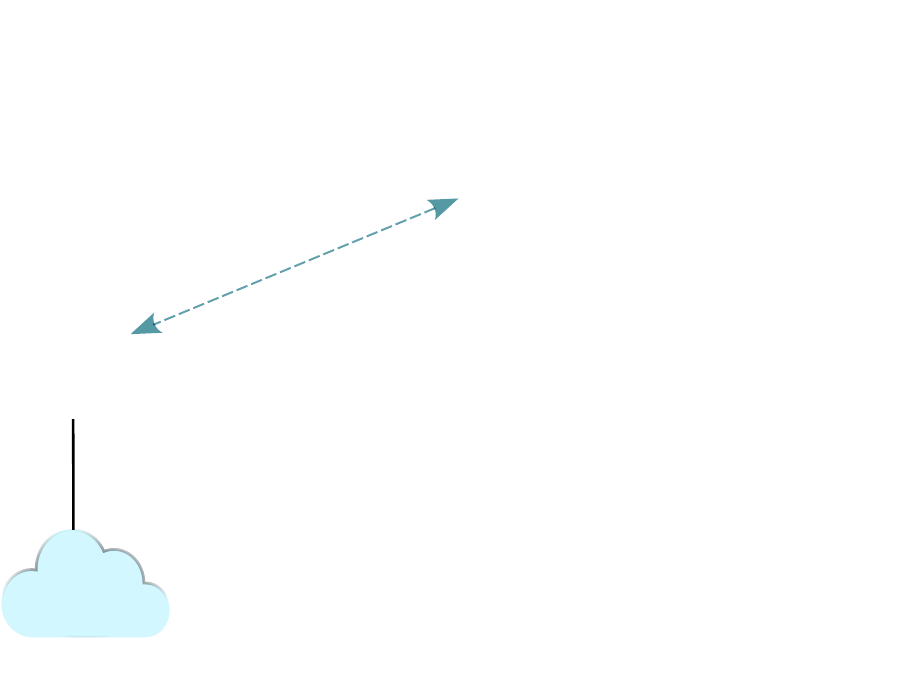
\caption{Schematic diagram of the downlink massive MIMO-enabled HAPS system with user geometry.}\label{fig:system_model}
\end{figure} 

\section{System Model}\label{sec:system_model}
We consider the downlink of a HAPS-enabled massive MIMO system in which a stratospheric platform at altitude \(H\) serves \(K\) randomly distributed single-antenna users within its coverage footprint, as illustrated in Fig.~\ref{fig:system_model}.
Let \(r_k\) denote the horizontal distance between the ground projection of the HAPS and user \(k\). The corresponding slant range is given by 
\begin{equation}
    d_k = \sqrt{H^2 + r_k^2}.
\end{equation}
The elevation angle for user \(k\), measured from the ground plane, is given by
\begin{equation}
    \vartheta_k = \arctan\!\left(\frac{H}{r_k}\right).
\end{equation}
For array response modeling, the corresponding angle relative to the array broadside (i.e., the downward-facing normal of the HAPS array) is
   $\theta_k = \frac{\pi}{2} - \vartheta_k.$
The HAPS is equipped with a uniform planar array (UPA) with \(M_x\) and \(M_y\) antenna elements along the two planar dimensions, where \(M=M_xM_y\). 

\subsection{Channel Model}
The downlink channel from HAPS to user \(k\) is modeled as a Rician fading channel. The instantaneous channel vector of user \(k\) is given by
\begin{equation}
\mathbf{h}_k
=
\sqrt{\beta_k}
\left(
\sqrt{\frac{\kappa_k}{\kappa_k+1}}\,\mathbf{a}(\theta_k,\varphi_k)
+
\sqrt{\frac{1}{\kappa_k+1}}\,\mathbf{R}_k^{1/2}\mathbf{z}_k
\right),
\label{eq:channel_model}
\end{equation}
where \(\beta_k\) is the large-scale channel gain, and \(\kappa_k\) is the Rician \(K\)-factor.  \(\mathbf{a}(\theta_k,\varphi_k)\in\mathbb{C}^{M\times 1}\) denotes the LoS steering vector, where \(\theta_k\) and \(\varphi_k\) denote the broadside and azimuth angles of departure (AoD), respectively. The matrix \(\mathbf{R}_k\in\mathbb{C}^{M\times M}\) captures the spatial correlation of the scattered component. Furthermore, \(\mathbf{z}_k \sim \mathcal{CN}(\mathbf{0},\mathbf{I}_M)\) is a complex Gaussian random vector with independent and identically distributed (i.i.d.) entries.
The correlation matrix \(\mathbf{R}_k\) is normalized such that $\mathrm{tr}(\mathbf{R}_k)=M$,
ensuring that \(\beta_k\) represents the average large-scale channel gain and that the Rician factor \(\kappa_k\) preserves its interpretation as the LoS-to-scattered power ratio.
The large-scale gain \(\beta_k\) is obtained from the 3GPP TR~38.811 NTN models~\cite{3gppTR38811V1540} as
\begin{equation}
    \beta_k = 10^{-\mathrm{PL}_k/10},
\end{equation}
where \(\mathrm{PL}_k\) is the total path loss (in dB), given by
\begin{equation}
    \mathrm{PL}_k = \mathrm{PL}_{b,k} + \mathrm{PL}_{g,k} + \mathrm{PL}_{s,k}.
\end{equation}
where \(\mathrm{PL}_{b,k}\) is the basic path loss, \(\mathrm{PL}_{g,k}\) accounts for atmospheric gas attenuation, and \(\mathrm{PL}_{s,k}\) models tropospheric scintillation. The basic path loss is
\begin{equation}
    \mathrm{PL}_{b,k} = \mathrm{PL}_{\mathrm{FS}}(d_k,f_c) + \mathrm{SF}_k + \mathrm{CL}_k,
\end{equation}
where  \(\mathrm{PL}_{\mathrm{FS}}(d_k, f_c)\) is the free-space path loss at slant range \(d_k\) (in kilometers) and carrier frequency \(f_c\) (in GHz). \(\mathrm{SF}_k \sim \mathcal{N}(0,\sigma_{\mathrm{SF}}^2)\) denotes shadow fading, and \(\mathrm{CL}_k\) represents clutter loss. The free-space path loss is expressed as
\begin{equation}
    \mathrm{PL}_{\mathrm{FS}}(d_k,f_c)
    =
    92.45 + 20\mathrm{log}_{10}(f_c) + 20\mathrm{log}_{10}(d_k).
\end{equation}
For the UPA, the LoS steering vector is
\begin{equation}
    \mathbf{a}(\theta_k,\varphi_k)
    =
    \mathbf{a}_x(\theta_k,\varphi_k)\otimes \mathbf{a}_y(\theta_k,\varphi_k),
\end{equation}
with \(\|\mathbf{a}(\theta_k,\varphi_k)\|^2=M\). Here, \(\mathbf{a}_x(\theta_k,\varphi_k)\) and \(\mathbf{a}_y(\theta_k,\varphi_k)\) are given by \vspace{-2mm}
\begin{subequations}
\begin{align}
\mathbf{a}_x(\theta_k,\varphi_k)
&=
\Big[
1,\,
e^{-j2\pi \frac{d_x}{\lambda}\sin\theta_k\cos\varphi_k},\, \notag\\
&\qquad \dots, 
e^{-j2\pi (M_x-1)\frac{d_x}{\lambda}\sin\theta_k\cos\varphi_k}
\Big]^T, \\
\mathbf{a}_y(\theta_k,\varphi_k)
&=
\Big[
1,\,
e^{-j2\pi \frac{d_y}{\lambda}\sin\theta_k\sin\varphi_k},\,  \notag\\
&\qquad \dots,
e^{-j2\pi (M_y-1)\frac{d_y}{\lambda}\sin\theta_k\sin\varphi_k}
\Big]^T.
\end{align}
\end{subequations}
\noindent where \(d_i\) for \(i \in \{x,y\} \) is the antenna spacing, and $\lambda$ is carrier wavelength. The first- and second-order statistics of \(\mathbf{h}_k\) are
\begin{subequations}
\begin{align}
    \bar{\mathbf{h}}_k
    &=
    \mathbb{E}[\mathbf{h}_k]
    =
    \sqrt{\beta_k}\sqrt{\frac{\kappa_k}{\kappa_k+1}}\,
    \mathbf{a}(\theta_k,\varphi_k), \\
    \mathbf{C}_k
    &=
    \mathrm{Cov}(\mathbf{h}_k)
    =
    \beta_k\frac{1}{\kappa_k+1}\mathbf{R}_k. 
\end{align}
\end{subequations} 

\subsection{Channel State Information (CSI)}
Acquiring accurate instantaneous CSI at the HAPS is impractical due to the large antenna dimension and wide-area coverage. In frequency division duplexing (FDD) operation, the feedback overhead scales with the number of antennas \(M\), rendering full instantaneous CSI acquisition prohibitively difficult \cite{Ibanez2025}. 
Moreover, the broad coverage footprint leads to a large and spatially diverse user population, thereby increasing pilot overhead and feedback load.
In contrast, statistical CSI, namely \(\bar{\mathbf{h}}_k\), \(\mathbf{C}_k\), and large-scale fading parameters, evolves on a timescale that is orders of magnitude slower than the instantaneous channel, governed by user mobility and shadow fading coherence rather than fast fading. This slow variation makes statistical CSI reliably estimable with low pilot overhead and robust to feedback latency \cite{You2020}. Further, for HAPS channels with a dominant LoS component, the statistical parameters are largely determined by geometry (user location, elevation angle, array response), making them relatively stable and predictable. Therefore, we assume that the HAPS employs statistical CSI-based precoding as the basis for beamformer design.

\subsection{Signal Model and Rate Analysis}
Let \(\mathbf{s}=[s_1,\dots,s_K]^T\) denote the transmit symbol vector satisfying \(\mathbb{E}[\mathbf{s}\mathbf{s}^H]=\mathbf{I}_K\). The HAPS downlink transmit signal is given by \vspace{-2mm}
\begin{equation}
    \mathbf{x} = \sum_{k=1}^K \sqrt{p_k}\,\mathbf{w}_k s_k,
\end{equation}
where \(p_k \ge 0\) is the downlink transmit power allocated to user \(k\), satisfying \(\sum_{k=1}^K p_k \le P_{\max}\), with \(P_{\max}\) denoting the maximum allowable transmit power at the HAPS. Moreover, \(\mathbf{w}_k \in\mathbb{C}^{M\times 1}\) is the precoding vector. Under statistical CSI, normalized mean-channel MRT is adopted. Since the scalar factors in \(\bar{\mathbf h}_k\) do not affect the normalized beam direction, the resulting precoder is \vspace{-2mm}
\begin{equation} \label{eq:precoder_w}
    \mathbf{w}_k = \frac{\mathbf{a}(\theta_k,\varphi_k)}{\sqrt{M}}.
\end{equation}

\noindent The received signal at user \(k\) can be written as
\begin{equation}
    y_k =
    \sqrt{p_k}\mathbf{h}_k^H\mathbf{w}_k s_k
    + \sum_{j\neq k}\sqrt{p_j}\mathbf{h}_k^H\mathbf{w}_j s_j
    + n_k,
\end{equation}
where \(n_k \sim \mathcal{CN}(0,\sigma^2)\) is the additive white Gaussian noise. 

Using the use-and-then-forget (UATF) bound~\cite{Massive_MIMO_BOOK_2016}, the achievable rate of user \(k\) is given by
\begin{equation}
    R_k = \mathrm{log}_2(1+\gamma_k),
\end{equation}
where \(\gamma_k\) is the effective SINR of user \(k\), given by
\begin{equation}\label{eq:effective_SINR}
\gamma_k =
\frac{p_k \left| \mathbb{E}\left[\mathbf{h}_k^H \mathbf{w}_k \right] \right|^2 }
{\sum_{j=1}^{K} p_j \mathbb{E}\left[\left| \mathbf{h}_k^H \mathbf{w}_j \right|^2 \right] 
- p_k \left| \mathbb{E}\left[\mathbf{h}_k^H \mathbf{w}_k \right] \right|^2  + \sigma^2}.
\end{equation}
By substituting (\ref{eq:channel_model}) and  (\ref{eq:precoder_w}) into (\ref{eq:effective_SINR}) and evaluating the resulting expectations, the effective SINR of user $k$ can be expressed in closed form as
\vspace{-4mm}

\begin{small}
    \begin{equation}
\!\!\gamma_k \!=\!
\frac{
p_k \beta_k \dfrac{\kappa_k}{\kappa_k\!+\!1} M
}{
\dfrac{\beta_k}{M(\kappa_k\!+\!1)}
\sum\limits_{j=1}^K p_j \mathbf{a}_j^H \mathbf{R}_k \mathbf{a}_j
+
\beta_k \dfrac{\kappa_k}{\kappa_k\!+\!1}
\sum\limits_{j\neq k} p_j \dfrac{|\mathbf{a}_k^H \mathbf{a}_j|^2}{M}
\!+\!
\sigma^2
}.
\label{eq:final_sinr}
\end{equation}
\end{small}

\noindent where \(\mathbf a_i \triangleq \mathbf a(\theta_i,\varphi_i)\) for \(i \in \{1,\dots,K\} \). The first term in the denominator captures the non-coherent scattered contribution. Under the UATF bound, this term includes both the variance of the desired signal (corresponding to $j=k$) and the scattered multiuser interference (corresponding to $j\neq k$), as all non-coherent components are treated as effective noise. The second term represents coherent LoS interference due to non-orthogonal user steering directions, highlighting the geometry-dependent nature of HAPS downlink transmission.

\section{Proposed Downlink Power Control}
Building on the SINR expression in \eqref{eq:final_sinr}, we develop downlink power-control policies for the considered HAPS massive MIMO system. The resulting interference structure induces a user-coupling model governed jointly by channel statistics and HAPS-user geometry.
To expose this structure, we define the effective desired-signal gain as
\begin{equation}
    g_k \triangleq \beta_k \frac{\kappa_k}{\kappa_k+1} M,
\end{equation}
the self-variance coefficient as
\begin{equation}
    c_{k,k} \triangleq \frac{\beta_k}{M(\kappa_k+1)} \mathbf a_k^H \mathbf R_k \mathbf a_k,
\end{equation}
and the inter-user coupling coefficients as
\begin{equation}
    c_{k,j} \triangleq
    \frac{\beta_k}{M(\kappa_k+1)} \mathbf a_j^H \mathbf R_k \mathbf a_j
    +
    \beta_k \frac{\kappa_k}{\kappa_k+1}\frac{|\mathbf a_k^H \mathbf a_j|^2}{M},
    \qquad j\neq k,
\end{equation}
Then, the SINR of user \(k\) can be rewritten as
\begin{equation}
    \gamma_k(\mathbf p)
    =
    \frac{g_k p_k}{c_{k,k}p_k + \sum_{j\neq k} c_{k,j}p_j + \sigma^2},
    \label{eq:sinr_compact}
\end{equation}
where \(\mathbf p=[p_1,\dots,p_K]^T \succeq \mathbf 0\) is the transmit power allocation vector.
The SINR expression in \eqref{eq:sinr_compact} reveals that HAPS downlink power control depends on both large-scale channel gains and geometry-induced spatial coupling. Based on this compact form, we next develop two power-control strategies tailored to different operational objectives.

\subsection{QoS-Constrained Minimum-Power Allocation} \label{subsec:qos_powercontrol}
We first consider the problem of minimizing the total downlink transmit power while guaranteeing per-user SINR targets. Let \(\bar{\gamma}_k = 2^{R_k^{\mathrm{req}}}-1\) denote the target SINR corresponding to the required rate \(R_k^{\mathrm{req}}\) of user \(k\). The QoS-constrained power-control problem is formulated as
\begin{equation}
\begin{aligned}
    \min_{\mathbf p \succeq \mathbf 0}\quad & \mathbf 1^T\mathbf p \\
    \text{subject to}\quad & \gamma_k(\mathbf p)\ge \bar{\gamma}_k,\quad \forall k,
\end{aligned}
\label{eq:qos_problem}
\end{equation}

\noindent where \(\mathbf 1\) denotes the all-ones vector. Substituting \eqref{eq:sinr_compact} into the constraints of \eqref{eq:qos_problem} yields
\begin{equation}
    \left(g_k-\bar{\gamma}_k c_{k,k}\right)p_k
    \ge
    \bar{\gamma}_k\sum_{j\neq k} c_{k,j}p_j + \bar{\gamma}_k \sigma^2.
\label{eq:qos_rearranged}
\end{equation}

\noindent A necessary condition for feasibility is
\begin{equation}
    g_k > \bar{\gamma}_k c_{k,k},\qquad \forall k.
\label{eq:qos_feasibility_basic}
\end{equation}

\noindent
This condition represents a user-specific feasibility limit under statistical CSI. In particular, if the self-variance coefficient \(c_{k,k}\) is large relative to the mean-channel gain \(g_k\), then user \(k\) cannot support the target SINR \(\bar{\gamma}_k\), even in the absence of multiuser interference. Equation \eqref{eq:qos_rearranged} can be rewritten in a standard interference-coupling form by defining
\begin{equation} \label{eq:eta_define}
    \eta_k \triangleq \frac{\bar{\gamma}_k}{g_k-\bar{\gamma}_k c_{k,k}},\qquad \forall k,
\end{equation}

\noindent which is positive under feasibility condition in  \eqref{eq:qos_feasibility_basic}, ensuring that \(\eta_k > 0\). Then, by substituting \eqref{eq:eta_define} into \eqref{eq:qos_rearranged}, we have
\begin{equation}
    p_k \ge \eta_k \left(\sum\nolimits_{j\neq k} c_{k,j}p_j + \sigma^2\right),\qquad \forall k.
\label{eq:qos_fp_scalar}
\end{equation}

\noindent Based on \eqref{eq:qos_fp_scalar}, we define the vector mapping \(F:\mathbb{R}_+^K \to \mathbb{R}_+^K\) as
\begin{equation}
    [F(\mathbf p)]_k
    \triangleq
    \eta_k \left(\sum\nolimits_{j\neq k} c_{k,j}p_j + \sigma^2\right),\qquad \forall k.
\end{equation}
Then, the QoS constraints can be written as the vector inequality
\begin{equation}
    \mathbf p \succeq F(\mathbf p).
\label{eq:qos_vector_fp}
\end{equation}

\noindent To obtain a matrix representation, we define the diagonal matrix \(\mathbf D=\mathrm{diag}(\eta_1,\dots,\eta_K)\), and the nonnegative coupling matrix \(\mathbf Z \in \mathbb{R}_+^{K\times K}\) with entries
\begin{equation}
    [\mathbf Z]_{k,j}=
    \begin{cases}
        c_{k,j}, & j\neq k,\\
        0, & j=k.
    \end{cases}
\end{equation}
\noindent Thus, the mapping \(F(\mathbf p)\) can be written in matrix form as
\begin{equation}
    F(\mathbf p) = \mathbf D \mathbf Z \mathbf p + \mathbf D \sigma^2 \mathbf 1.
\end{equation}

\noindent Hence, the minimum feasible power vector satisfies the fixed point equation
\begin{equation}
    \mathbf p = \mathbf D \mathbf Z \mathbf p + \mathbf D \sigma^2 \mathbf 1.
\label{eq:qos_matrix_fp}
\end{equation}

\noindent
\textbf{Proposition 1:}
\textit{Under condition \eqref{eq:qos_feasibility_basic}, the QoS problem in \eqref{eq:qos_problem} is feasible if and only if}
    $\rho(\mathbf D \mathbf Z) < 1$.
\textit{In that case, the unique minimum-power solution is}
\begin{equation}
    \mathbf p^\star = (\mathbf I - \mathbf D \mathbf Z)^{-1}\mathbf D \sigma^2 \mathbf 1.
    \label{eq:qos_closed_form}
\end{equation}

\textit{Proof:}
Let \(\mathbf A=\mathbf D\mathbf Z\) and
\(\mathbf b=\mathbf D\sigma^2\mathbf 1\succ\mathbf 0\).
The QoS constraints are equivalent to
\(\mathbf p\succeq\mathbf A\mathbf p+\mathbf b\).
For the nonnegative matrix \(\mathbf A\), a finite nonnegative feasible power vector exists if and only if
\(\rho(\mathbf A)<1\). Under this condition,
\(\mathbf I-\mathbf A\) is a nonsingular M-matrix and $
(\mathbf I-\mathbf A)^{-1}
=\sum_{n=0}^{\infty}\mathbf A^n\succeq\mathbf 0$ 
\cite{BermanPlemmons1994}. Hence, the unique fixed point is
\(\mathbf p^\star=(\mathbf I-\mathbf A)^{-1}\mathbf b\).
Moreover, any feasible \(\mathbf p\) satisfies
\(\mathbf p\succeq\mathbf p^\star\), so \(\mathbf p^\star\) is componentwise minimal and therefore minimizes
\(\mathbf 1^T\mathbf p\).
\hfill\(\blacksquare\)


While \eqref{eq:qos_closed_form} provides a closed-form solution, it requires matrix inversion, which may be impractical for large systems or real-time implementation. Alternatively, the solution can be obtained via a low-complexity fixed-point iteration based on the mapping \(F(\mathbf p)\), which converges to \(\mathbf p^\star\) whenever \(\rho(\mathbf D\mathbf Z)<1\). The procedure is summarized in Algorithm~\ref{alg:qos_fp}.

\begin{algorithm}[t]
\caption{QoS Power Control via Fixed-Point Iteration}
\label{alg:qos_fp}
\KwIn{\(\bar{\gamma}_k\), tolerance \(\epsilon\), max iterations \(N_{\max}\)}
\KwOut{\(\mathbf p^\star\) or infeasibility}

Compute \(g_k, c_{k,k}, c_{k,j}\)\;

\For{each \(k\)}
{
    \If{\(g_k \le \bar{\gamma}_k c_{k,k}\)}{return infeasible}
    \(\eta_k = \bar{\gamma}_k/(g_k-\bar{\gamma}_k c_{k,k})\)\;
}

Construct \(\mathbf D, \mathbf Z\)\;

\If{\(\rho(\mathbf D\mathbf Z)\ge 1\)}{return infeasible}

Initialize \(\mathbf p^{(0)}=\mathbf 0\)\;

\For{\(n=0\) to \(N_{\max}-1\)}
{
    Update $p_k^{(n+1)}=\eta_k\left(\sum\nolimits_{j\neq k}c_{k,j}p_j^{(n)}+\sigma^2\right)$\\
    \If{converged}{return \(\mathbf p^\star\)}
}
Return non-convergence warning\;
\end{algorithm}

\subsection{Max--Min Fairness Power Allocation}
In contrast to the power minimization problem considered in Section~\ref{subsec:qos_powercontrol}, where the objective is to meet given SINR targets with minimum transmit power, here we consider a max–min fairness formulation that aims to maximize the minimum achievable SINR among all users under a total transmit power budget. The problem is formulated as 
\begin{equation}
\begin{aligned}
    \max_{\mathbf p \succeq 0,\; t}\quad & t\\
    \text{subject to}\quad & \gamma_k(\mathbf p)\ge t,\quad \forall k,\\
    & \mathbf 1^T\mathbf p \le P_{\max}.
\end{aligned}
\label{eq:maxmin_problem}
\end{equation}
where $t$ denotes the common SINR target for all users. By following steps similar to those in \eqref{eq:qos_rearranged}–\eqref{eq:qos_closed_form}, the minimum power vector to support a given SINR target \(t\) is given by
\begin{equation}
    \mathbf p^\star(t)=
    (\mathbf I-\mathbf \Phi(t)\mathbf Z)^{-1}\mathbf \Phi(t)\sigma^2 \mathbf 1, \label{eq:p_t_star}
\end{equation}
\noindent where \(\mathbf \Phi(t)=\mathrm{diag}(\varphi_k(t))\), with \( \varphi_k(t)=t\big/\big(g_k-t c_{k,k}\big)\).
The  feasibility of \(t\) requires
\begin{equation}\label{eq:maxmin_feasibilty}
    g_k>t c_{k,k},\quad \rho(\mathbf \Phi(t)\mathbf Z)<1,\quad
    \mathbf 1^T\mathbf p^\star(t)\le P_{\max}.
\end{equation}

Similar to Algorithm~\ref{alg:qos_fp}, for each fixed \(t\), the power vector in \eqref{eq:p_t_star} can also be obtained via fixed-point iteration, which converges to \(\mathbf p^\star(t)\) whenever $\rho(\mathbf\Phi(t)\mathbf Z)<1$. This shows that max--min fairness reduces to finding the largest SINR level \(t\) that satisfies both the user-specific feasibility condition and the global power constraint.
A practical upper bound on \(t\) is obtained by considering the interference-free case in which all transmit power is allocated to a single user, yielding
\begin{equation}
    t_{\max} = \min_k \frac{g_k P_{\max}}{c_{k,k}P_{\max}+\sigma^2}.
\end{equation}

\noindent The optimal value \(t^*\)
 can be computed by performing bisection search over the interval \([0, t_{\max}]\), where at each iteration the feasibility of a candidate \(t\) is verified using \eqref{eq:p_t_star}--\eqref{eq:maxmin_feasibilty} by checking the conditions in \eqref{eq:maxmin_feasibilty}. This bisection-based procedure is summarized in Algorithm~\ref{alg:maxmin}.
\begin{algorithm}[t]
\caption{Max--Min Fairness via Bisection}
\label{alg:maxmin}
\KwIn{\(P_{\max}, \epsilon\)}
\KwOut{\(t^\star, \mathbf p^\star\)}

Initialize \(t_{\mathrm{low}}=0\), \(t_{\mathrm{high}}=t_{\max}\)\;

\While{\(t_{\mathrm{high}}-t_{\mathrm{low}}>\epsilon\)}
{
    \(t=(t_{\mathrm{low}}+t_{\mathrm{high}})/2\)\;

    Compute \(\mathbf p^\star(t)\) and check feasibility conditions\;

    \If{feasible}{
        \(t_{\mathrm{low}}=t\), store \(\mathbf p^\star\)
    }
    \Else{
        \(t_{\mathrm{high}}=t\)
    }
}
\Return \(t^\star=t_{\mathrm{low}}, \mathbf p^\star\)\;
\end{algorithm}

\subsection{Benchmark Schemes}\label{sub_sec_benchmarks}
We compare the proposed power-control schemes against the following baseline strategies.

\subsubsection{Uniform Power Allocation} 
This scheme allocates equal power to all users, without accounting for path-loss disparity or spatial interference coupling:
\[
p_k = \frac{P_{\max}}{K}.
\]

\subsubsection{Inverse Path-Loss Allocation} 
This approach compensates for the near-far effect by favoring users with weaker large-scale channel gains, but does not account for inter-user interference:
\[
p_k = \frac{P_{\max}\beta_k^{-1}}{\sum_i \beta_i^{-1}}.
\]

\subsubsection{Geometry-Aware Heuristic}
This heuristic allocates power proportional to the aggregate interference-to-gain ratio, thereby prioritizing users with unfavorable channel conditions. It serves as a fairness-oriented baseline that compensates for geometry-induced disparities while incorporating spatial coupling effects:
\[
p_k \propto \frac{c_{k,k}+\sum_{j\neq k}c_{k,j}}{g_k}.
\]

\subsubsection{LP-Based Benchmark Methods}
For comparison purposes, the QoS-constrained problem in \eqref{eq:qos_problem} is also solved as a linear program using CVX \cite{Grant2020}. Similarly, for the max--min problem in \eqref{eq:maxmin_problem}, a bisection search is performed over the common SINR target \(t\), and, for each fixed value of \(t\), the corresponding linear feasibility problem is solved using CVX.

\section{Numerical Results}\label{sec_results}
We evaluate the proposed schemes for a HAPS at \(H=20~\mathrm{km}\) serving users randomly distributed within a circular footprint of radius \(20~\mathrm{km}\) using a \(10\times10\) UPA (\(M=100\)). The carrier frequency and bandwidth are \(2.4~\mathrm{GHz}\) and \(10~\mathrm{MHz}\), respectively, with \(\lambda/2\) antenna spacing and a \(7~\mathrm{dB}\) noise figure. The shadow-fading standard deviation and Rician \(K\)-factor are \(4~\mathrm{dB}\) and \(10~\mathrm{dB}\), respectively, while clutter, atmospheric, and scintillation losses are \(0\), \(0.3\), and \(0.5~\mathrm{dB}\). \(\mathbf{R}_k\) is modeled using an exponential correlation model with coefficient 0.7. Unless otherwise stated, results are averaged over \(N_{\mathrm{MC}}=2500\) independent deployments. Simulations were conducted in MATLAB R2025a on an Intel Core Ultra 9 275HX CPU with 32 GB RAM.



\begin{figure}[t!]\centering 
	\includegraphics[width=0.35\textwidth]{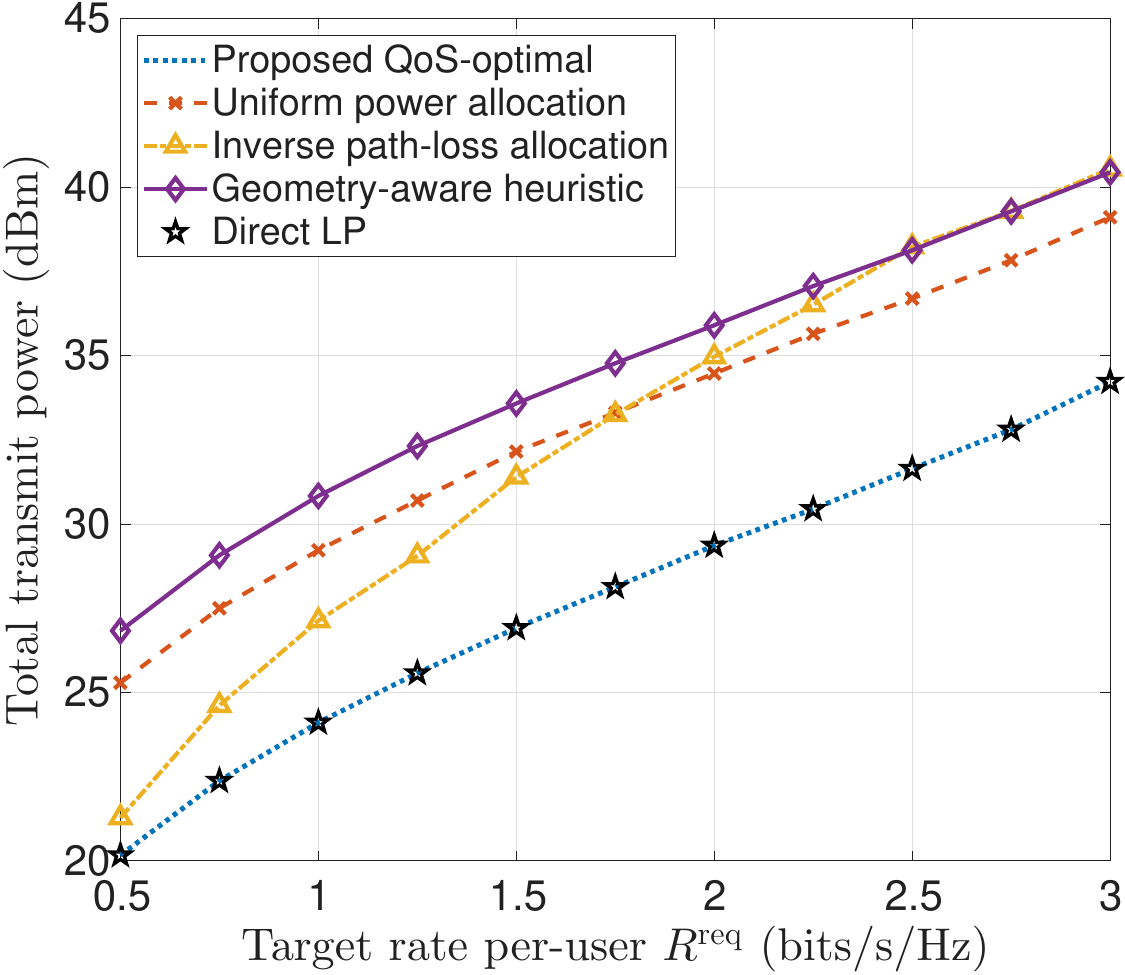}\vspace{-2mm}
	\caption{Required total transmit power versus target per-user rate under QoS-Constrained power allocation for $K=10$.}\label{fig:QosProblem_vs_Rate}\vspace{-2mm}
\end{figure}

Fig.~\ref{fig:QosProblem_vs_Rate} shows the required transmit power versus the target per-user rate \(R^\text{req}\). The proposed QoS-optimal scheme requires substantially less power than the uniform, inverse path-loss, and geometry-aware heuristic allocations across all target rates, while closely approaching the Direct LP benchmark. The increasing gap of the inverse path-loss scheme at higher rates highlights the importance of accounting for inter-user interference coupling. The geometry-aware heuristic requires higher power because its fairness-oriented allocation does not directly minimize the power required to satisfy the QoS constraints. These results confirm that incorporating geometry-aware interference modeling, including both scattered and LoS components, is essential for energy-efficient power control in HAPS systems.


\begin{figure}
  \centering
  \includegraphics[width=0.35\textwidth]{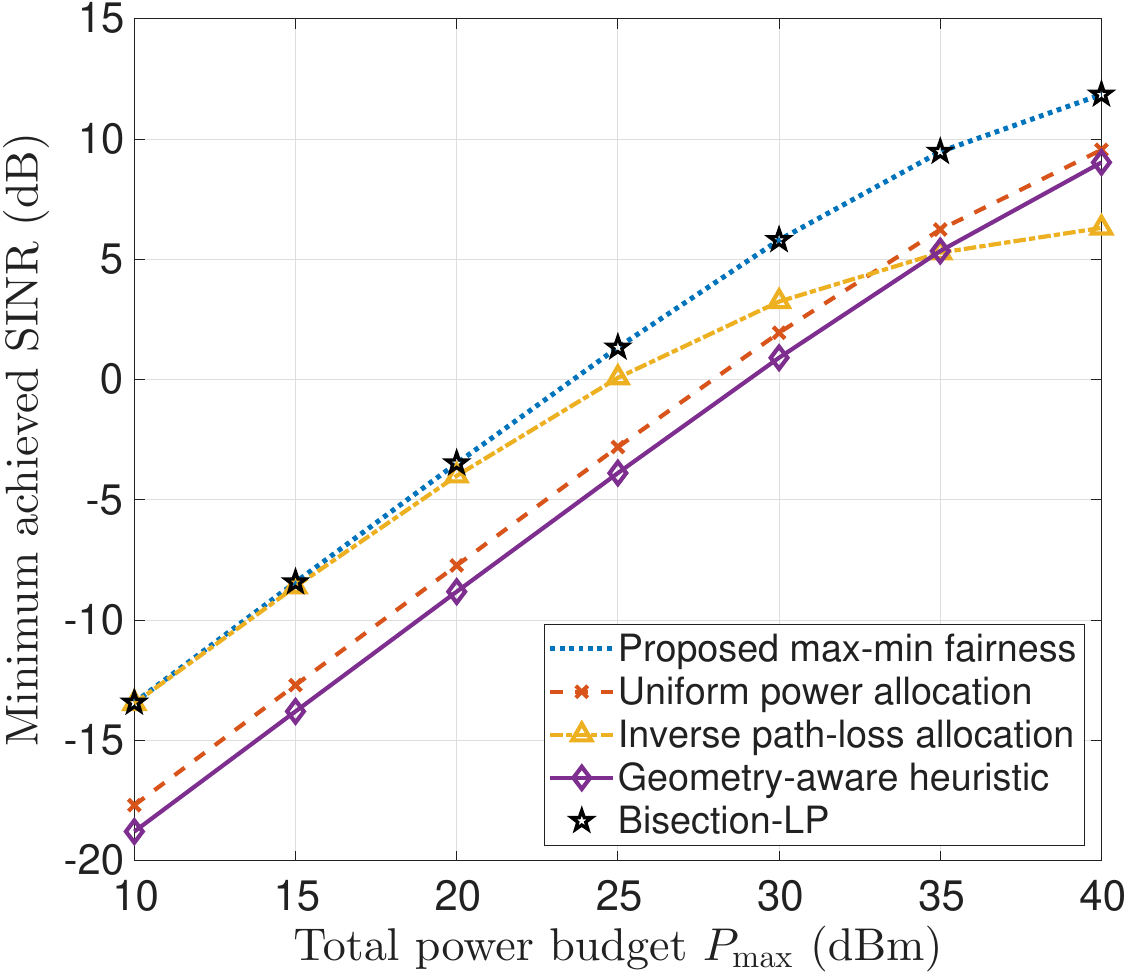}\vspace{-2mm}
   \caption{Minimum achieved SINR versus total transmit power under max–min fairness for $K=10$.}\vspace{-4mm}    \label{fig:max-min} 
\end{figure}

Fig.~\ref{fig:max-min} shows the minimum achieved SINR versus the total power budget \(P_{\max}\). The proposed max--min scheme closely approaches the Bisection-LP benchmark and consistently outperforms the uniform, inverse path-loss, and geometry-aware heuristic allocations. The inverse path-loss scheme performs closest at low-to-moderate \(P_{\max}\) by partially compensating for path-loss disparities, while the geometry-aware heuristic performs worse due to its mismatch with the actual interference coupling structure. These results confirm that interference-aware power control is essential for fairness across diverse users in HAPS systems.


\begin{figure}
  \centering
  \includegraphics[width=0.35\textwidth]{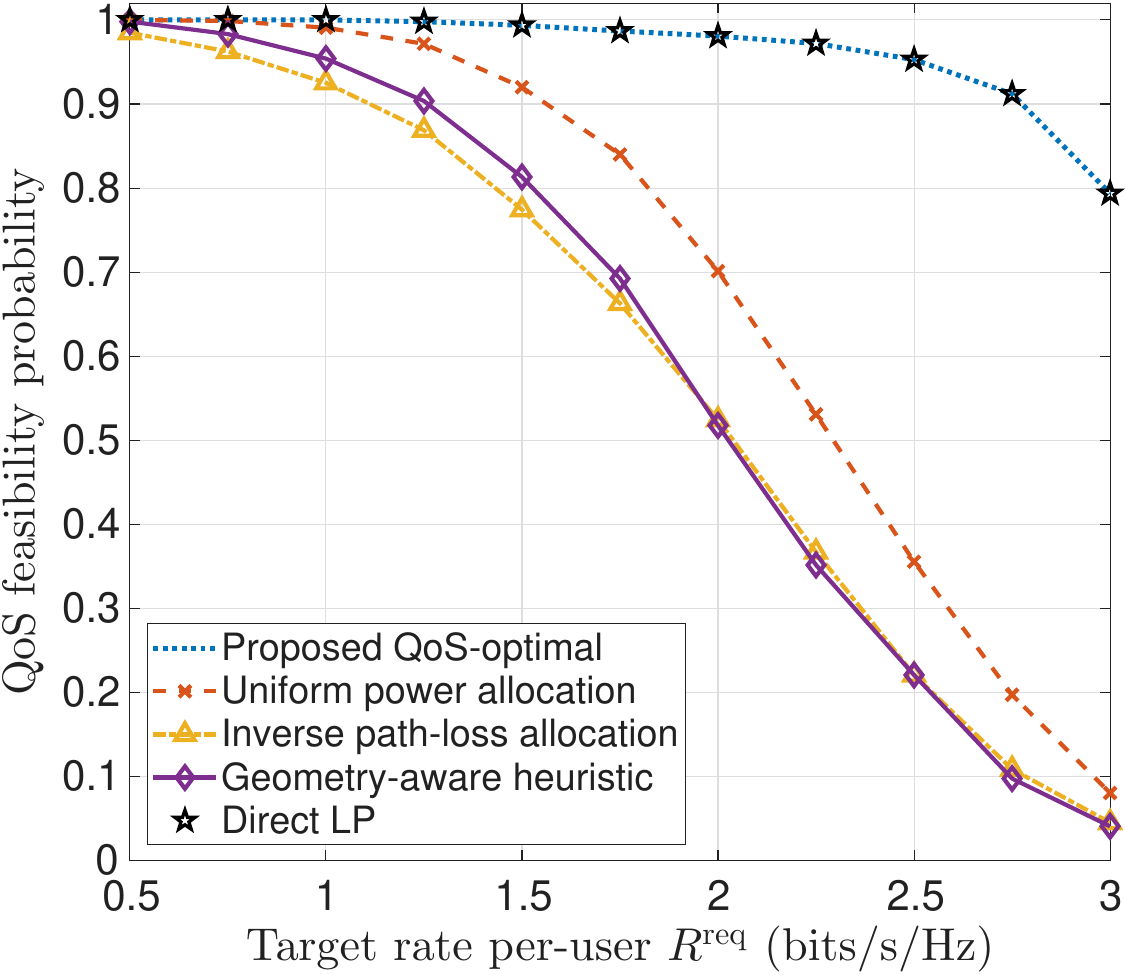}\vspace{-2mm}
   \caption{QoS feasibility probability versus target per-user rate for $K=10$, and $P_{\max} =35$ dBm.}   \label{fig:Qos_vs_Rate} 
\end{figure}

\begin{figure}
  \centering
  \includegraphics[width=0.35\textwidth]{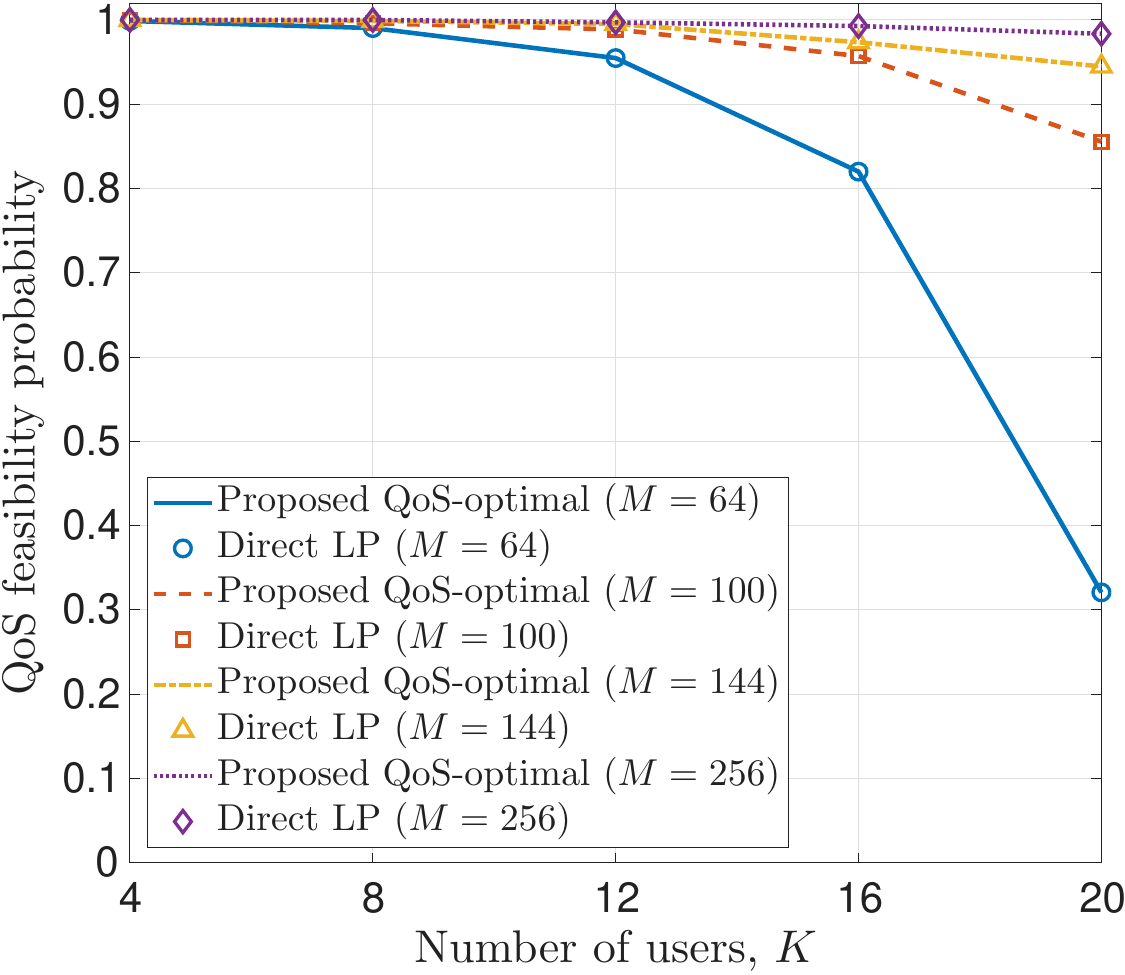}\vspace{-2mm}
   \caption{QoS feasibility probability versus number of users for different number of HAPS antennas, with $R^\text{req} = 1.5$ bit/s/Hz and $P_{\max} =35$ dBm.}\vspace{-2mm}    \label{fig:Qos_vs_K} 
\end{figure}

To characterize QoS performance under the total power budget, we define the QoS feasibility probability as\vspace{-2mm}
\begin{equation}
\label{eq:Qos_feas}\vspace{-1mm}
P_{\mathrm{QoS}}=\frac{N_\mathrm{feasible}}{N_\mathrm{MC}},
\end{equation}
where \(N_\mathrm{feasible}\) is the number of realizations in which all users satisfy \(R_k\ge R^{\mathrm{req}}\) and \(\mathbf 1^T\mathbf p\le P_{\max}\).

Fig.~\ref{fig:Qos_vs_Rate} shows that the proposed QoS-optimal scheme achieves feasibility comparable to Direct LP and substantially higher than the baseline allocations, particularly at high target rates. At \(R^\text{req}=2.5\) bits/s/Hz, its feasibility probability is approximately \(0.95\), compared with less than \(0.4\) for all three baselines. Fig.~\ref{fig:Qos_vs_K} further shows that feasibility decreases with \(K\) and improves with \(M\), with the proposed scheme remaining comparable to Direct LP.



\begin{figure}[t]
  \centering
  \includegraphics[width=0.46\textwidth]{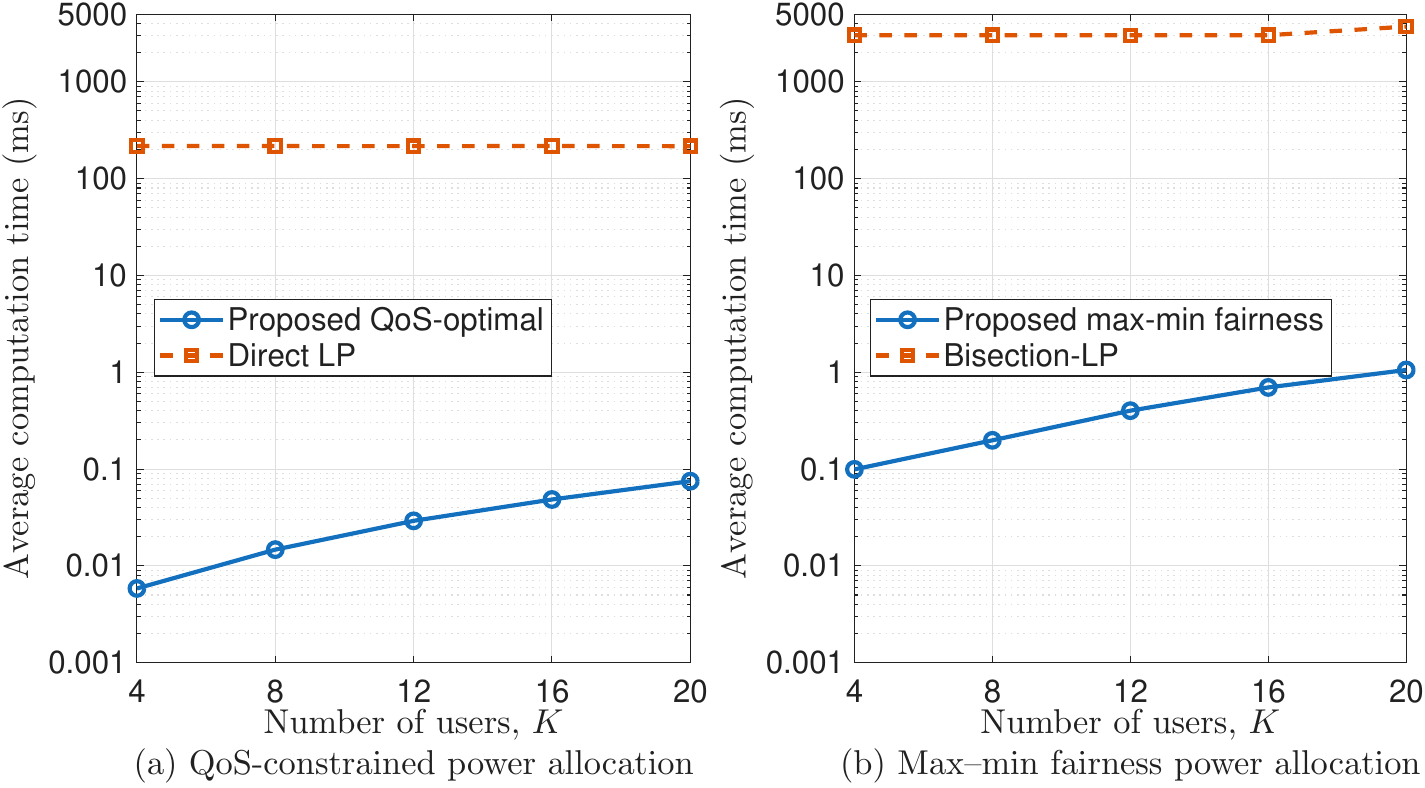}\vspace{-2mm}
   \caption{Average computation time versus number of users, with $R^\text{req} = 1.5$ bit/s/Hz and $P_{\max} =35$ dBm.}\vspace{-6mm}    \label{fig:computation_time} 
\end{figure}
Fig.~\ref{fig:computation_time} compares the computation time of the proposed schemes and LP-based benchmarks over 100 independent realizations. At \(K=20\), both proposed schemes reduce computation time by approximately \(99.97\%\). By avoiding LP-based optimization, these reductions support power control on computationally constrained HAPS platforms.


\section{Conclusion}

This paper developed a downlink power-control framework for HAPS massive MIMO under statistical CSI and 3GPP NTN channel models. A closed-form SINR expression under mean-channel MRT was derived, capturing both scattered and geometry-dependent coherent LoS interference and enabling tractable solutions to the QoS-constrained minimum-power and max--min fairness problems. Numerical results demonstrated lower transmit power, improved minimum-user SINR, and higher QoS feasibility than conventional power allocations. Moreover, the proposed schemes achieved performance comparable to LP-based benchmarks but with much lower computational complexity, reducing computation time by approximately $99.97\%$ at $K=20$. These results highlight the potential of the proposed low-complexity power-control framework for efficient and reliable HAPS deployment and operation.

\linespread{0.98}
	\bibliographystyle{IEEEtran}
	\bibliography{IEEEabrv,References}
	
\end{document}